\documentclass{article}
\usepackage{spconf,amsmath,graphicx,comment}
\usepackage{amssymb} 
\usepackage{multirow}
\usepackage{CJKutf8}
\newcommand*{\jpn}[1]{\begin{CJK}{UTF8}{ipxm}{\small #1}\end{CJK}}
\usepackage[multiple]{footmisc}

\usepackage{color}
\usepackage{tikz}

\usepackage[hyperfootnotes=false]{hyperref}
\usepackage{booktabs}

\title{Polyphone disambiguation and accent prediction using pre-trained language models in Japanese TTS front-end}
\name{Rem Hida$^{\star}$, Masaki Hamada$^{\star}$, Chie Kamada$^{\dagger}$, Emiru Tsunoo$^{\star}$, Toshiyuki Sekiya$^{\star}$, Toshiyuki Kumakura$^{\dagger}$}
\address{$^{\star}$ Sony Group Corporation, Tokyo, Japan \qquad
$^{\dagger}$ Sony Corporation of America, San Jose, CA, USA
}
\begin{document}
\ninept
\maketitle

\begin{abstract}
Although end-to-end text-to-speech (TTS) models can generate natural speech, challenges still remain when it comes to estimating sentence-level phonetic and prosodic information from raw text in Japanese TTS systems. 
In this paper, we propose a method for polyphone disambiguation (PD) and accent prediction (AP). The proposed method incorporates explicit features extracted from morphological analysis and implicit features extracted from pre-trained language models (PLMs).
We use BERT and Flair embeddings as implicit features and examine how to combine them with explicit features.
Our objective evaluation results showed that the proposed method improved the accuracy by 5.7 points in PD and 6.0 points in AP. 
Moreover, the perceptual listening test results confirmed that a TTS system employing our proposed model as a front-end achieved a mean opinion score close to that of synthesized speech with ground-truth pronunciation and accent in terms of naturalness.
\end{abstract}

\vspace{-0.3cm}
\section{Introduction}\label{sec:intro}
\vspace{-0.15cm}

The quality of text-to-speech (TTS) systems has improved in recent years to approach human levels of naturalness, owing to the development of deep learning-based approaches~\cite{Shen18taco2,ren2021fastspeech}.
Traditional TTS systems consist of three parts: a TTS front-end, an acoustic model, and a vocoder.
The TTS front-end extracts linguistic features, including phonetic and prosodic features from raw text, the acoustic model converts linguistic features into acoustic features such as a mel-spectrogram, and the vocoder produces waveforms from acoustic features.
One of the difficulties in TTS systems is the high language dependency of the front-end.
Recent studies have shown that end-to-end TTS systems, especially Japanese ones, require not only phonetic but also prosodic information in order to achieve high quality speech~\cite{Fujimoto2019,Kiyoshi}.
To enable the TTS front-end to precisely estimate both phonetic and prosodic information, two problems must be solved: polyphone disambiguation (PD) and accent prediction (AP).
PD estimates the correct pronunciation of polyphonic words. 
Polyphonic words have multiple candidate pronunciations, and the correct pronunciation depends on the context.

AP consists of accent phrase boundary prediction (APBP) and accent nucleus position prediction (ANPP). 
APBP chunks words into accent phrases, and ANPP determines where pitches change from high to low in each accent phrase.

\begin{figure*}[t]
    \centering
    \includegraphics[height=6cm]{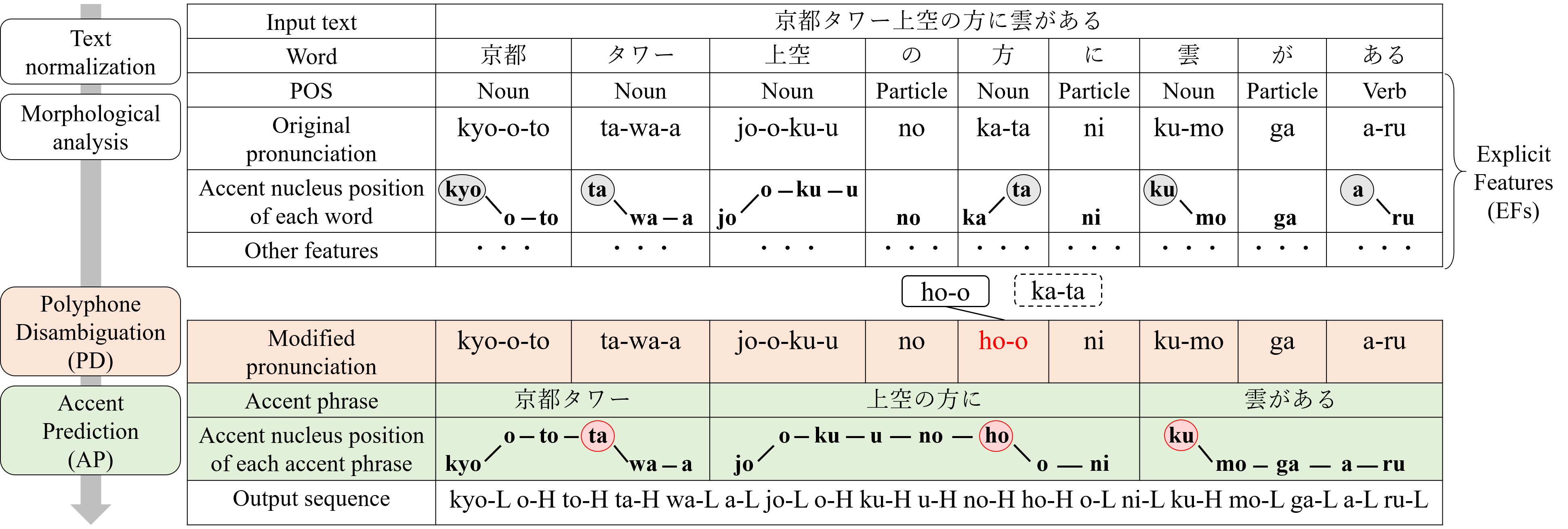}
    \vspace{-0.5cm}
    \caption{\small Pipeline of TTS front-end for Japanese}
    \label{fig:pipeline_tts}
    \vspace{-0.3cm}
\end{figure*}

In this paper, we propose the application of pre-trained language models (PLMs) for performing PD and AP. 
PLMs have been successfully used for Japanese phrase break prediction~\cite{futamata2021phrase}, for TTS front-ends in other languages~\cite{Bai21,talman-etal-2019-predicting,Yang2019}, and as additional input features for acoustic models~\cite{Hayashi2019,Kenter2020,Zack21,jia2021png}.
We combine the explicit features derived from morphological analysis and the implicit features derived from PLMs.
Explicit features include part-of-speech (POS), pronunciation, accent type, and other kinds of linguistic information. 
These features contain phonetic and prosodic information that cannot be obtained from raw text.
On the other hand, thanks to their bi-directional architectures and massive training corpora, PLMs such as BERT and Flair can provide context-aware information that cannot be obtained from morphological analysis.  
Thus, explicit features and implicit features are complementary to each other.

Our contributions are summarized as follows:
\begin{enumerate}
    \item To the best of our knowledge, this work is the first to incorporate PLMs for PD and AP in a Japanese TTS front-end. 
    \item We investigated strategies for combining explicit linguistic features with PLMs.
    \item Our proposed method improves both PD and AP performance in objective evaluations.
    \item In the subjective listening test, the proposed method achieved almost the same quality as synthesized speech with ground-truth pronunciation and accent.
\end{enumerate}

\begin{table*}[t]
    \centering
    \small
    \vspace{-0.2cm}
    \caption{Description of explicit features (EFs). The right three columns show the EFs used in each task.}
    \label{table:explicit_features}
    \footnotesize
    \begin{tabular}{l@{\hspace{2mm}}ll|ll@{\hspace{2mm}}l@{\hspace{2mm}}}
        Feature ID & Feature type &  Description &EF$_{\text{PD}}$&EF$_{\text{APBP}}$&EF$_{\text{ANPP}}$  \\ \hline\hline
        EF1&POS& Part-of-speech tag&\checkmark&\checkmark&\checkmark\\
        EF2&Basic linguistic features&Form and type of conjugation, word type, etc.   &&\checkmark&\checkmark\\
        EF3&Phonetic features & The number of morae, the first two morae in the pronunciation, etc. &&\checkmark&\checkmark \\
        EF4&Prosodic features (word) & Accent type, accent combination type, etc. &&\checkmark&\checkmark  \\
        EF5&Prosodic features (accent phrase) & Number of words in accent phrase, etc. & & & \checkmark\\
        EF6&Accent nucleus position rule features   & How an accent nucleus change according to rules & & & \checkmark \\
        EF7&$n$-gram features  & Unigram and bigram information derived from Wikipedia &&(Optional)&\\
    \end{tabular}
    \vspace{-0.55cm}
\end{table*}
\vspace{-0.3cm}
\section{Problem setting}
\vspace{-0.15cm}

The Japanese TTS front-end aims to convert the input text into  linguistic features, including phonetic and prosodic features.
Fig.~\ref{fig:pipeline_tts} shows a pipeline of the TTS front-end for an input text ``\jpn{京都タワー上空の方に雲がある} (There are clouds above Kyoto Tower)." This consists of four components: text normalization, morphological analysis, PD, and AP. 
The input text is first normalized and then tokenized by a morphological analyzer, which also assigns linguistic features to each word. 
Subsequently, words with linguistic features are input into the PD and AP modules.
Finally, they are converted into a sequence of pronunciations with the pitch (low or high) of each mora.
In the following sections, we describe the problem setting of PD and AP.
We also briefly explain the conventional approaches to these problems.

\vspace{-0.3cm}
\subsection{Polyphone disambiguation (PD)} 
\vspace{-0.15cm}

Kanji characters in Japanese have multiple candidate pronunciations, and each pronunciation corresponds to a different meaning.
For example, in Fig.~\ref{fig:pipeline_tts}, ``\jpn{方}" can be pronounced in two ways with two respective meanings: ``ho-o" (direction) and ``ka-ta" (way or person). 
Conventional morphological analysis still fails to estimate the pronunciation of such polyphonic words, because it uses only local context.
Therefore, an additional PD module that can consider meaning in the context is necessary for estimating the pronunciation of polyphonic words.

Some studies have explored solving PD by regarding pronunciation estimation (including non-polyphonic words) as a sequence-to-sequence problem, and by applying machine translation approaches~\cite{hatori2011japanese,kakegawa21_interspeech}.
For Mandarin, on the other hand, some studies adopt a classification approach that estimates the correct pinyin of the polyphonic character~\cite{Park2020chinese, Sun2019polyphone, Cai2019polyphone}.
Because polyphonic words appear only in certain parts of the sentence, we regard PD as a classification problem, similar to the approach for Mandarin.

\vspace{-0.3cm}
\subsection{Accent prediction (AP)}
\vspace{-0.15cm}
Japanese is a pitch-accent language that has accent phrases and accent nucleus positions.
The accent phrase is a unit in which a pitch ascent occurs, followed by a descent.
The accent nucleus position is the mora just before the pitch descends in the accent phrase.
These features are not explicitly written in Japanese raw text; however, they are important for prosodic naturalness in Japanese TTS systems~\cite{Fujimoto2019,Kiyoshi}. 

AP consists of two parts: APBP and ANPP.
APBP estimates whether each word boundary is an accent phrase boundary in order to chunk words into accent phrases.
ANPP estimates the accent nucleus position change of each word in each accent phrase.
We regard APBP and ANPP as sequence labeling problems, because the correct accent labels depend on the context of the label sequence.

APBP is particularly challenging in the cases of adjacent nouns.
For example, there is no boundary between the first adjacent nouns ``\jpn{京都} (Kyoto)" and ``\jpn{タワー} (Tower)" in Fig.~\ref{fig:pipeline_tts}, but there is one between the second adjacent nouns ``\jpn{タワー}" and ``\jpn{上空} (above)."
Accent nucleus positions also change depending on the context. 
For example, Fig.~\ref{fig:pipeline_tts} shows the accent nucleus position change for ``\jpn{京都}". Its original accent nucleus position is the first mora ``kyo," as indicated by the circle in Fig.~\ref{fig:pipeline_tts}.
However, its accent nucleus position changes when it is compounded by the following word ``\jpn{タワー}."
To address these challenges, several approaches have been proposed for AP,
such as conventional rule-based methods~\cite{sagisaka1983accentuation}, statistical methods~\cite{suzuki2017accent}, and deep learning methods~\cite{kakegawa21_interspeech}.

\vspace{-0.3cm}
\section{Proposed Method}\label{sec:proposed_method}
\vspace{-0.15cm}
        \begin{table*}[!ht]
            \centering
            \caption{Estimation examples of BiLSTM models for PD}
            \label{table:result_PD_example}
        \begin{tabular}{c|ccc}
            Example&EF$_{\text{PD}}$ (PD3)& EF$_{\text{PD}}$ + BERT (PD6) &Reference            \\ \hline\hline
            \begin{tabular}[c]{@{}c@{}}\jpn{今は}\underline{\jpn{辛い}}\jpn{過去も忘れて幸せに暮らしている。}\\ (Now I forget the \textbf{painful} past and live happily.)\end{tabular}&``ka-ra-i" (spicy)
            & {\bf ``tsu-ra-i" (painful)}& ``tsu-ra-i" (painful) \\\hline
            \begin{tabular}[c]{@{}c@{}} \jpn{試合の前日は}\underline{\jpn{辛い}}\jpn{ものを作ることが少なくありません。}\\ (It is not uncommon to make \textbf{spicy} food the day before a game.)\end{tabular}&``tsu-ra-i" (painful)&{\bf ``ka-ra-i" (spicy)} & ``ka-ra-i" (spicy)  
        \end{tabular}
        \vspace{-0.5cm}
        \end{table*}
The proposed model combines the explicit features derived from morphological analysis and the implicit features derived from PLMs.
As stated previously, explicit features contain not only basic linguistic information but also phonetic and prosodic information, whereas implicit features can represent contextualized information. Thus, explicit and implicit features are complementary.

\vspace{-0.3cm}
\subsection{Explicit features}
\vspace{-0.15cm}
\label{subsec:explicit linguistic features}

We categorize the explicit features (EFs) into seven parts in Table~\ref{table:explicit_features}. 
EF1--4 are derived from morphological analysis.
EF5 is derived from APBP, EF6 is derived from accent nucleus position rules, and EF7 is derived from text corpora such as Wikipedia.
The three right-side columns of Table~\ref{table:explicit_features} show the EFs used in each task.
Since PD is based on the meaning of words, only EF1 is used.
APBP and ANPP are related to phonetic and prosodic phenomena; thus, in addition to EF1, EF2--4 are used for both, EF5 and 6 are used for ANPP, and EF7 is optionally used for APBP, based on~\cite{suzuki2017accent}.
In summary, EF$_{\text{PD}}$, EF$_{\text{APBP}}$, and EF$_{\text{ANPP}}$ denote EF1, EF1--4, and EF1--6, respectively.

\vspace{-0.3cm}
\subsection{Implicit features}
\vspace{-0.15cm}
\label{subsec:PLM}
In this paper, we propose the use of two types of PLMs as implicit features: BERT and Flair.
BERT~\cite{devlin2019bert} is a Transformer~\cite{vaswani2017attention} encoder that is pre-trained by using masked language modeling and next sentence prediction objectives on a raw text corpus.
Flair~\cite{akbik-etal-2018-contextual} is an LSTM~\cite{Hochreiter1997lstm} encoder that is pre-trained by using a character-level language modeling objective on a raw text corpus.
Thanks to their bi-directional architectures and massive training corpora, both of them have the capability to capture long context and semantic information.
One of the differences between BERT and Flair is the token unit.
BERT encodes text by subword, and Flair encodes text by character.
Additionally, previous studies have shown that BERT contains some syntactic information~\cite{Hewitt2019probe} and can distinguish the sense of words more clearly than Flair~\cite{Wiedemann2019sense}.
Based on this, we hypothesize that BERT is more suitable for PD and APBP because they are related to syntactic and semantic information, and Flair is more suitable for ANPP, which is related to finer units of information such as phonemes and accents.

\vspace{-0.3cm}
\subsection{Proposed model with explicit and implicit features}
\vspace{-0.15cm}
\label{subsec:proposed_model}

\begin{figure}[t]
    \centering
    \includegraphics[width=8.8cm]{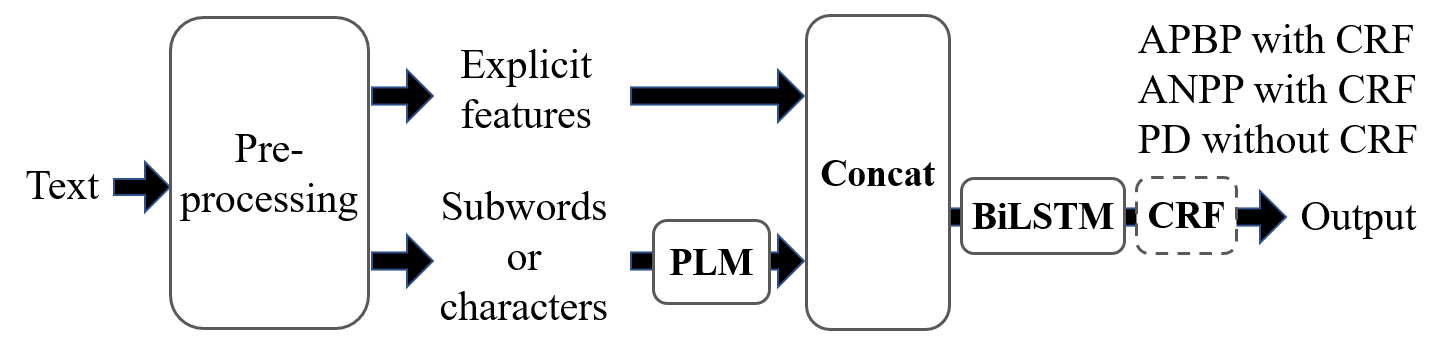}
    \caption{Proposed model architecture}
    \vspace{-0.55cm}
    \label{fig:model_architecture}
\end{figure}

Fig.~\ref{fig:model_architecture} shows the architecture of the proposed model.
The text is pre-processed by using the morphological analyzer MeCab\mbox{}\footnote{\url{https://taku910.github.io/mecab/}} with the dictionary UniDic\mbox{}\footnote{\url{https://unidic.ninjal.ac.jp/}}, which performs tokenization and POS tagging and extracts other information explicitly.
A PLM extracts linguistic information implicitly from token sequences.
Subsequently, the embeddings of the explicit features and the PLM are concatenated and input into the BiLSTM layer. For PD, the pronunciation of the polyphonic word is the output. For APBP and ANPP, since they are sequence labeling problems, the hidden states of the BiLSTM layer are then passed through an additional conditional random field (CRF)~\cite{Lafferty2001crf} layer to output the accent phrase boundary and accent nucleus position. The outputs of APBP are binary labels which indicate whether there is an accent phrase boundary before each word, and the outputs of ANPP are multiple labels which describe how the original accent nucleus positions change, based on \cite{suzuki2017accent}.

\vspace{-0.3cm}
\section{Experiments}\label{sec:experiments}
\vspace{-0.15cm}
    \subsection{Polyphone disambiguation (PD) objective evaluation}
    \subsubsection{Experimental setup}
    \vspace{-0.1cm}
    \label{sec:experiments_PD_setup}
    We compared the performance of the proposed method with baselines as an objective evaluation.
    We collected 39,353 sentences as an in-house dataset sampled from Wikipedia, TV captions, novels, CSJ~\cite{maekawa2003csj}, and JSUT~\cite{sonobe2017jsut} and manually annotated the pronunciation in each case. The dataset included 39,897 polyphonic words.
    We split the dataset into training, validation, and test sets of 24,117, 5,156, and 10,080 sentences, respectively.
    The JNAS corpus~\cite{jnas} was also used as a public test set.
    In the experiments, we focused on 92 frequently used polyphonic words.

    Our baselines were based on morphological analysis.
    We used MeCab and a KyTea-trained model for comparison.
    The KyTea-trained model is a morphological analyzer that uses pointwise prediction, and it was trained using the aforementioned training data, based on~\cite{neubig2010word}.
    The proposed method is based on BiLSTM, explained in Sec.~\ref{subsec:proposed_model}.
    We adopted a one-layer BiLSTM (with 512 units), which was implemented with a Flair framework~\cite{akbik2019flair}. The model was trained using an SGD optimizer with a mini-batch size of 32. 
    The initial learning rate was 0.1, which was halved each time the validation accuracy did not improve for four consecutive epochs.
    We stopped training when the learning rate fell below $10^{-4}$.
    As implicit features, the BERT-base model and the Flair model which are pre-trained on Japanese Wikipedia were used\mbox{}\footnote{\url{https://huggingface.co/cl-tohoku/bert-base-japanese-v2}}\footnote{``ja-forward" and ``ja-backward" on \url{https://github.com/flairNLP/flair}} .
    When BERT was used as an implicit feature, the last four layers were concatenated\mbox{}\footnote{Preliminary experiments showed that using all the layers performed as well or worse than using only the last four layers.}.
    During the training, the parameters of the PLM were fixed\mbox{}\footnote{Preliminary  experiments showed that there was no significant improvement achieved by fine-tuning the PLM.}.
    The BiLSTM models were trained with seven different random seeds, and the average accuracy of each test set was reported.

    \vspace{-0.3cm}
    \subsubsection{Experimental results}
    \vspace{-0.15cm}
    Table~\ref{table:result_PD} shows the accuracy of the baselines and BiLSTM models for PD. 
    In both the in-house and JNAS test sets, the BiLSTM model with both explicit and implicit features achieved the highest performance.
    In the in-house data, the BiLSTM model with EF$_{\text{PD}}$ and BERT (PD6) improved accuracy by 5.7 points over PD2, which was the best model without implicit features.
    We found that BERT was generally superior to Flair as an implicit feature.
    This is in line with the same findings in the word sense disambiguation task~\cite{Wiedemann2019sense}, which is similar to PD in terms of requiring semantic information.
    
    Table~\ref{table:result_PD_example} shows the examples of PD in different BiLSTM systems.
    The proposed method using EF$_{\text{PD}}$ and BERT correctly predicted the pronunciation, while the BiLSTM with only EF$_{\text{PD}}$ failed to predict these examples.
    These examples indicate that BERT contributes to taking context into account and improving PD performance.
        \begin{table}[t]
        \vspace{-0.3cm}
            \centering
            \caption{Performance of different systems on PD in the in-house (IH) dataset and JNAS. PD1 and 2 use a morphological analyzer only. EF$_{\text{PD}}$ denotes the POS.
            * denotes a statistically significant difference with the best performance at p $<.05$.}
            \label{table:result_PD}
            \scalebox{0.95}{
                \begin{tabular}{c|c@{\hspace{1mm}}c@{\hspace{3mm}}c|c}
                ID & Model& Explicit& Implicit & \begin{tabular}[c]{@{}c@{}}Accuracy\\ (IH/JNAS)\end{tabular} \\ \hline\hline
                PD1&MeCab&---&---& 81.03*/96.27*\\
                PD2&KyTea-trained~\cite{neubig2010word} &---&---& 88.67*/96.63*\\
                PD3&BiLSTM& EF$_{\text{PD}}$&---& 88.15*/95.24* \\
                PD4&BiLSTM&---& BERT& 94.24/96.64\\
                PD5&BiLSTM&---& Flair& 92.86*/96.00*\\ \hline
                PD6&BiLSTM& EF$_{\text{PD}}$ & BERT& \textbf{94.34}/96.72 \\
                PD7&BiLSTM& EF$_{\text{PD}}$ & Flair& 93.46*/\textbf{96.77}\\
                \end{tabular}
            }
            \vspace{-0.55cm}
        \end{table}

    \vspace{-0.3cm}
    \subsection{Accent prediction (AP) objective evaluation}
    \subsubsection{Experimental setup}
    \vspace{-0.15cm}
    As an objective evaluation, we compared the performance in APBP, ANPP, and overall AP. We collected 9,497 sentences as an in-house dataset sampled from TV captions. 
    We split the dataset into training, validation, and test sets of 7,768, 864, and 865 sentences, respectively.
    We also used a sub-corpus (basic5000) of JSUT~\cite{sonobe2017jsut} and its accent label\mbox{}\footnote{\url{https://github.com/sarulab-speech/jsut-label}} as a public test set for overall AP performance.
    The basic5000 test set consists of 5,000 sentences along with their pronunciations. 
    Sentences with pronunciations that did not match any of the 5-best analysis results from MeCab were excluded.
    As a result, 4,210 sentences were left for evaluation.
    
    We compared the combination of explicit features (EFs, Sec.~\ref{subsec:explicit linguistic features}) and implicit features (PLMs, Sec.~\ref{subsec:PLM}) in the proposed method, the rule-based method~\cite{sagisaka1983accentuation}, and TASET\mbox{}\footnote{\url{https://sites.google.com/site/suzukimasayuki/accent}}.
    The TASET is an accent prediction tool that uses CRF~\cite{suzuki2017accent}, and it was trained on the same training set.
    For the BiLSTM experiments, we adopted an architecture and learning setting similar to PD. The only difference was the addition of a CRF layer.
    We trained the model with five different random seeds and reported the average of the F1-score or accuracy on a test set.
    For the overall AP evaluation, we compared pairs of different APBP and ANPP models in terms of accuracy.
    \vspace{-0.3cm}
    \subsubsection{Accent phrase boundary prediction (APBP) results}
    \vspace{-0.15cm}
        Table~\ref{table:result_APBP} shows the F1-score for all words and adjacent nouns.
        The results show that the systems using EF$_{\text{APBP}}$ and BERT achieved the highest F1-scores.
        We observed the following two findings. 
        First, incorporating PLMs as implicit features was effective for APBP, where BERT was better than Flair, especially for adjacent nouns.
        Second, $n$-gram features were effective for adjacent nouns even when Flair was adopted, but not when BERT was adopted.
        This implies that the BERT embedding provides $n$-gram information.
    \vspace{-0.3cm}
    \subsubsection{Accent nucleus position prediction (ANPP) results}
    \vspace{-0.15cm}
        Table~\ref{table:result_ANPP} shows the accuracy for all accent phrases and a subset of long accent phrases having more than two words.
        The results show that the system using EF$_{\text{ANPP}}$ and Flair embeddings achieved the highest accuracy.
        We observed that EF$_{\text{ANPP}}$ was a powerful feature for ANPP, when compared with only PLMs. 
        The results also show that there is room for improvement in the model using only EF$_{\text{ANPP}}$, compared to the model using both EF$_{\text{ANPP}}$ and Flair.
        The results indicate that Flair is more effective than BERT for ANPP, which requires finer unit information such as phonemes and accents.
        
        \begin{table}[t]
            \centering
            \caption{F1-score of different systems on APBP. 
                    EF$_{\text{APBP}}$ denotes the explicit features for APBP (see Sec.~\ref{subsec:explicit linguistic features}).}
            \label{table:result_APBP}
            \scalebox{0.95}{
                \begin{tabular}{c@{\hspace{3mm}}c|c@{\hspace{3mm}}c}
                 Explicit  & Implicit &  All & Adjacent nouns\\\hline\hline
                \multicolumn{2}{c|}{Rule-based}  & 91.20*   &  47.00*      \\
                \multicolumn{2}{c|}{TASET~\cite{suzuki2017accent}}  & 95.43*   &  74.46*      \\\hline
                    EF$_{\text{APBP}}$ (+ $n$-gram)    &    ---       & 95.44* (95.53*)  & 77.94* (79.90*) \\
                    ---         &  BERT     & 95.95*  & 82.77*  \\
                    ---    &  Flair     & 95.91*   & 78.34*    \\\hline
                    EF$_{\text{APBP}}$ (+ $n$-gram)    &  BERT     & \textbf{96.30} (96.23)      &\textbf{85.61} (85.16)      \\
                    EF$_{\text{APBP}}$ (+ $n$-gram)   &  Flair     & 96.17 (96.18)   & 82.16* (85.26)         \\
                    
                \end{tabular}
            }
            \vspace{-0.55cm}
        \end{table}

        \begin{table}[t]
            \centering
            \caption{Accuracy of different systems on ANPP.
            EF$_{\text{ANPP}}$ denotes the explicit features for ANPP (see Sec.~\ref{subsec:explicit linguistic features}). 
            }
            \label{table:result_ANPP}
            \scalebox{0.95}{
                \begin{tabular}{cc|cc}
                Explicit & Implicit& All &Long accent phrases\\\hline\hline
                \multicolumn{2}{c|}{Rule-based}  & 83.07*   &  63.95*      \\
                \multicolumn{2}{c|}{TASET~\cite{suzuki2017accent}}  & 95.24*   &  90.83*      \\\hline
                EF$_{\text{ANPP}}$        & ---      & 95.34* &91.56*     \\
                ---            & BERT   & 84.38*   &78.63*          \\
                ---          & Flair  & 87.12*     &81.51*     \\\hline
                EF$_{\text{ANPP}}$        & BERT   & 94.57*  &90.19*         \\
                EF$_{\text{ANPP}}$  & Flair  &\textbf{95.99} &\textbf{92.98}
                \\\hline
                \end{tabular}
            }
            \vspace{-0.55cm}
        \end{table}
        
        \vspace{-0.3cm}
        \subsubsection{Overall accent prediction results}
        \vspace{-0.15cm}
        Table~\ref{table:result_APPipe} shows the exact match rate of the sentences (Snt-Exact) and the accuracy of the pitch of each mora (Mora-Accuracy) of each model.

        The pair of the APBP model using EF$_{\text{APBP}}$ + BERT and the ANPP model using EF$_{\text{ANPP}}$ + Flair, which were the best models for each task, outperformed the other pairs in both in-house and JSUT test sets.
        The in-house data showed that the best system improved over the model using only EFs by 6.0 points in terms of Snt-Exact and 0.5 points in terms of Mora-Accuracy.
        
        \begin{table}[t]
            \centering
            \caption{Performance of different systems on overall accent prediction in the in-house (IH) dataset and JSUT.
            }
            \label{table:result_APPipe}

            \scalebox{0.95}{
              \hspace{-0.5cm}                \begin{tabular}{@{\hspace{1mm}}c@{\hspace{1mm}}|@{\hspace{1mm}}c@{\hspace{2mm}}c@{\hspace{1mm}}|@{\hspace{1mm}}c@{\hspace{1mm}}c@{\hspace{1mm}}}
                    ID&APBP model&ANPP model& \begin{tabular}[c]{@{}c@{}}Snt-Exact\\(IH/JSUT)\end{tabular}&\begin{tabular}[c]{@{}c@{}}Mora-Accuracy\\(IH/JSUT)\end{tabular}\\\hline\hline
                    AP0 & Rule-based& Rule-based &17.34*/16.35* & 90.86*/96.56*\\
                    AP1 & EF$_{\text{APBP}}$ + $n$-gram & EF$_{\text{ANPP}}$ &52.67*/25.91*&96.15*/97.07*\\\hline
                    AP2& EF$_{\text{APBP}}$ + BERT & EF$_{\text{ANPP}}$ + Flair & \textbf{58.68}/\textbf{26.98}&\textbf{96.66}/\textbf{97.33} \\
                \end{tabular}
            }
            \vspace{-0.55cm}
        \end{table}
        
    \vspace{-0.3cm}
    \subsection{Text-to-speech quality subjective evaluation}
    \vspace{-0.15cm}
    
    To confirm the effectiveness of the proposed model in the Japanese TTS front-end, 
    we carried out a mean opinion score (MOS) test as a subjective evaluation.
    We adopted Tacotron2~\cite{Shen18taco2} with global style tokens~\cite{wang2018style} as an acoustic model and Parallel WaveGAN~\cite{Yamamoto2020parallel} as a vocoder. 
    We used the sub-corpus (basic5000) of JSUT and its accent label to train both the acoustic model and the vocoder.
    
    In the MOS test, 30 native Japanese speakers were asked to evaluate synthesized speech samples on a 5-point Likert scale.
    We compared the following four AP systems: 
    (1) \textbf{Oracle symbols}: annotated labels. 
    (2) \textbf{Rule-based}: a rule-based AP system (AP0 in Table~\ref{table:result_APPipe}). 
    (3) \textbf{EFs}: BiLSTM models using only EFs for both APBP and ANPP (AP1 in Table~\ref{table:result_APPipe}).
    (4) \textbf{Proposed}: our best combination of EF$_{\text{APBP}}$ + BERT APBP model and EF$_{\text{ANPP}}$ + Flair ANPP model (AP2 in Table~\ref{table:result_APPipe}). 
    For the MOS test, 25 utterances were randomly selected from the in-house test set of AP and were then synthesized with the aforementioned input variations from four systems. 
    We confirmed that the pronunciations of all 25 utterances were correctly estimated by our proposed PD model; thus, the effectiveness of AP was purely evaluated.
    
    Table~\ref{table:result_MOS} shows the MOS test results of different systems. It shows that the proposed method outperformed the system without PLMs (AP1) and achieved almost the same speech quality as oracle symbols. 
    We found that the subjective evaluation score was greatly influenced by the accuracy of the adjacent noun accents, which was especially improved by the proposed method.
    This indicates that using both explicit features and implicit PLM features is effective for the Japanese TTS in terms of naturalness.
    
    \begin{table}[t]
        \centering
        \caption{MOS evaluation of TTS with 95\% confidence intervals computed from the t-distribution for different systems.}
        \label{table:result_MOS}
        \scalebox{0.95}{
            \begin{tabular}{c|c}
                Systems  & MOS  \\ \hline\hline
                Oracle symbols & 3.69 $\pm$ 0.07            \\ \hline
                Proposed (AP2) & \textbf{3.67 $\pm$ 0.07}           \\
                EFs (AP1) & 3.46 $\pm$ 0.07              \\
                Rule-based (AP0) & 3.18 $\pm$ 0.08   \\
            \end{tabular}
        }
        \vspace{-0.55cm}
    \end{table}

\vspace{-0.3cm}
\section{Conclusion}
\vspace{-0.15cm}
In this paper, we demonstrated the effectiveness and characteristics of PLMs such as BERT and Flair for the Japanese TTS front-end. The objective evaluation results showed that the combination of explicit features from morphological analysis and implicit features from PLMs improved the performance of PD and AP compared with the individual implicit/explicit features.  
Moreover, the subjective evaluation results showed that our proposed method in the TTS front-end is effective for generating natural speech.

\vspace{-0.3cm}
\section{Acknowledgments}
\vspace{-0.15cm}
The authors would like to thank Takuma Okamoto at the National Institute of Information and Communications Technology for discussing the evaluation design.
\vfill\pagebreak

\bibliographystyle{IEEEbib}
\bibliography{strings,refs}

\end{document}